\documentclass{WileyMSP-template}
\usepackage{xcolor}
\usepackage[colorlinks]{hyperref}
\usepackage{siunitx}
\usepackage{amsmath}
\usepackage{booktabs}
\usepackage{float}
\usepackage[normalem]{ulem}

\usepackage{microtype}
\usepackage{ragged2e}

\microtypesetup{
    expansion=true,
    protrusion=true,
    tracking=true,
    kerning=true,
    spacing=true
}

\RaggedRight
\hypersetup{
    colorlinks=true,
    linkcolor=blue,
    urlcolor=blue,
    citecolor=blue,
}
\begin{document}

\pagestyle{fancy}
\rhead

\title{\textbf{Anomalous magneto-optical response at  RuO$_2$/WSe$_2$ van der Waals interface}}

\maketitle

\author{Muhammad Hassan Shaikh*},
\author{Abhijith Puthiya Veettil},
\author{Collin Maurtua},
\author{Dai Q. Ho},
\author{Subhash Bhatt},
\author{David T. Plouff},
\author{Malitha Gulawita},
\author{Kenji Watanabe},
\author{Takashi Taniguchi},
\author{John Q. Xiao},
\author{Anderson Janotti}, and
\author{Chitraleema Chakraborty*}.

\begin{affiliations}
M. H. Shaikh, S. Bhatt, D. T. Plouff, M. Gulawita, Dr. J. Q. Xiao, Dr. C. Chakraborty\\
Department of Physics, University of Delaware, Newark, Delaware 19716, USA.\\
Email: mhshaikh@udel.edu,  cchakrab@udel.edu\\
Phone: +1(302) 831-3251

A. P. Veettil, C. Maurtua, Dr. D.Q. Ho,  Dr. A. Janotti, Dr. C. Chakraborty\\
Department of Materials Science and Engineering, University of Delaware, Newark, Delaware 19716, USA.

Dr. K. Watanabe, Dr. T. Taniguchi\\
National Institute for Materials Science (NIMS), Tsukuba 305-0047, Japan.

Dr. J. Q. Xiao, Dr. C. Chakraborty\\
Quantum Science and Engineering Program, University of Delaware, Newark, Delaware 19716, USA.

\end{affiliations}


\keywords{Altermagnetism, van der Waals heterostructure, RuO$_2$, WSe$_2$, Surface magnetism, Polarization-resolved magneto-optical spectroscopy}


\begin{abstract}
\justifying{Ruthenium dioxide (RuO$_2$) has been proposed as an altermagnetic candidate, although its magnetic ground state remains controversial. Here, we probe weak interfacial magnetic states at the surface of (001)-oriented RuO$_2$ films using the magnetic proximity effect (MPE) in a van der Waals heterostructure consisting of monolayer tungsten diselenide (WSe$_2$) atop RuO$_2$. Temperature-dependent magneto-optical spectroscopy reveals an anomalous excitonic energy shift and a deviation from conventional Varshni behavior below 55 K that are absent in an encapsulated WSe$_2$ control sample. The anomalous shift reverses sign upon field cooling with opposite magnetic field polarity, indicating a magnetic origin. Polarization-resolved measurements further show a nearly field-independent and fluctuating valley splitting in WSe$_2$/RuO$_2$, in strong contrast to the conventional linear Zeeman splitting observed in the control bare WSe$_2$ sample. These results suggest that the valley states are governed predominantly by interfacial exchange fields associated with weak surface magnetic states in RuO$_2$, which do not produce a conventional linear Zeeman response within the applied magnetic field range. Importantly, this approach enables direct optical probing of emergent surface magnetism without introducing an additional ferromagnetic layer, positioning MPE-based optical probing as a tool for investigating weak surface magnetism and offering new possibilities for studying magnetic materials with controversial magnetic states.}

\end{abstract}


\justifying{
\section{Introduction}

Antiferromagnetic (AFM) materials are considered a promising solution for next-generation ultrafast spintronics and high-density memories due to their compensated magnetization and terahertz magnetic dynamics \cite{baltz_antiferromagnetic_2018, noauthor_antiferromagnetic_nodate, baltz_emerging_2024, dal_din_antiferromagnetic_2024, jungwirth_antiferromagnetic_2016, xiong_antiferromagnetic_2022, song_altermagnets_2025, bai_altermagnetism_2024, liu_antiferromagnetic_2025,jeong_altermagnetic_2026}. However, their lack of easily measurable physical observables related to their magnetic order parameter poses a significant challenge for utilizing these materials. Recently, a special class of antiferromagnetic materials called altermagnets has gained substantial attention \cite{song_altermagnets_2025, bai_altermagnetism_2024, liu_antiferromagnetic_2025}. In altermagnets, the opposite spin sublattices in real space are surrounded by an anisotropic crystal field and connected by a rotation symmetry, which leads to the emergence of spin-splitting band structures in momentum space \cite{song_altermagnets_2025, bai_altermagnetism_2024, liu_antiferromagnetic_2025, PhysRevX.12.040501}. This spin-splitting could give rise to direction-dependent pure spin currents or spin-polarized currents, which can be exploited for high-density memory and ultrafast spintronics applications.

Ruthenium dioxide (RuO$_2$) has been predicted to be an above-room-temperature altermagnetic candidate \cite{PhysRevX.12.040501}. However, there have been many contrasting reports regarding the magnetic ground state in RuO$_2$ films \cite{jeong_altermagnetic_2026,chi_crystal-facet-oriented_2024, PhysRevLett.132.166702, kesler_absence_2024, Kiefer_2025, zhang_simultaneous_2024, 6fxv-153y, plouff_revisiting_2025, abel2025probingmagneticpropertiesruo2, chen_giant_2026, akashdeep2026surface}. Muon spin rotation and neutron diffraction further revealed the absence of a magnetic ground state in bulk crystals, powder, and films of RuO$_2$ \cite{PhysRevLett.132.166702, kesler_absence_2024, Kiefer_2025}. Nevertheless, several experiments on RuO$_2$ and RuO$_2$-based heterostructures have revealed unusual interfacial exchange and spin-dependent phenomena at the surface \cite{abel2025probingmagneticpropertiesruo2, akashdeep2026surface}, including anomalous magnetic coupling behavior in RuO$_2$/ferromagnet bilayers \cite{abel2025probingmagneticpropertiesruo2}. Furthermore, a recent study by muon spin rotation experiment found an inhomogeneous magnetic state at the surface of RuO$_2$ films\cite{akashdeep2026surface}.
Together, these observations suggest the possible existence of interfacial or surface-localized electronic
states capable of generating surface magnetic moments or exchange fields without requiring bulk long-range
magnetic order.}

In this work, we employ monolayer tungsten diselenide (WSe$_2$), a transition metal dichalcogenide (TMDC) as an ultrasensitive optical probe of interfacial exchange interactions at the RuO$_2$ surface via the magnetic proximity effect (MPE). Unlike conventional proximity structures involving metallic ferromagnets such as NiFe or Fe \cite{abel2025probingmagneticpropertiesruo2}, the WSe$_2$/RuO$_2$ heterostructure enables direct interrogation of surface-induced exchange fields while minimizing magnetic proximity contributions from an adjacent ferromagnetic layer. MPE is a short-range interaction, and employing TMDC to study MPE-based effects has long been used to investigate magnetism in ferromagnets (FMs) and antiferromagnets (AFMs) \cite{zhao_interlayer_2024, zhong_van_2017, zhong_layer-resolved_2020, shaikh_magnetic_2025, chakraborty_flatland_2024, zhang_tuning_2026, beer_proximity-induced_2024}. Specifically, WSe$_2$ has been shown to be an excellent candidate for detecting surface-layer magnetic moments in both few-layer and bulk systems \cite{zhong_layer-resolved_2020, shaikh_magnetic_2025}. Magnetic field dependent and valley resolved photoluminescence (PL), therefore, provide a sensitive means to probe emergent surface magnetic states in RuO$_2$ thin films that may be inaccessible to conventional bulk magnetometry techniques.

We study two heterostructures: one comprising monolayer WSe$_2$ atop a RuO$_2$ (001) film (WSe$_2$/RuO$_2$) to detect surface layer magnetism, and a second heterostructure comprising monolayer WSe$_2$ encapsulated by hexagonal boron nitride (h-BN) as a control sample to decouple the influence of the RuO$_2$ film. Temperature-dependent magneto-optical spectroscopy of the WSe$_2$/RuO$_2$ heterostructure reveals an anomalous energy shift and deviation from conventional Varshni behavior below 55 K, indicating the emergence of interfacial exchange fields associated with a surface magnetic state in RuO$_2$. The sign of the anomalous shift reverses upon field cooling with opposite magnetic field polarity, indicating that the associated surface magnetization can be reoriented by the applied magnetic field. Furthermore, polarization-resolved measurements reveal markedly different valley-splitting behavior in WSe$_2$/RuO$_2$ compared with h-BN-encapsulated WSe$_2$, suggesting the presence of weak surface-induced exchange effects absent in the control sample. These observations support the existence of unconventional surface magnetic states in RuO$_2$. Our results establish WSe$_2$ as a highly sensitive optical probe for detecting surface exchange fields and emergent surface magnetism in complex oxide thin films that are difficult to access using conventional bulk magnetometry techniques.

\justifying{
\section{Results and discussion}
\begin{figure}[!ht]
    \centering
    \includegraphics[width=0.8\textwidth]{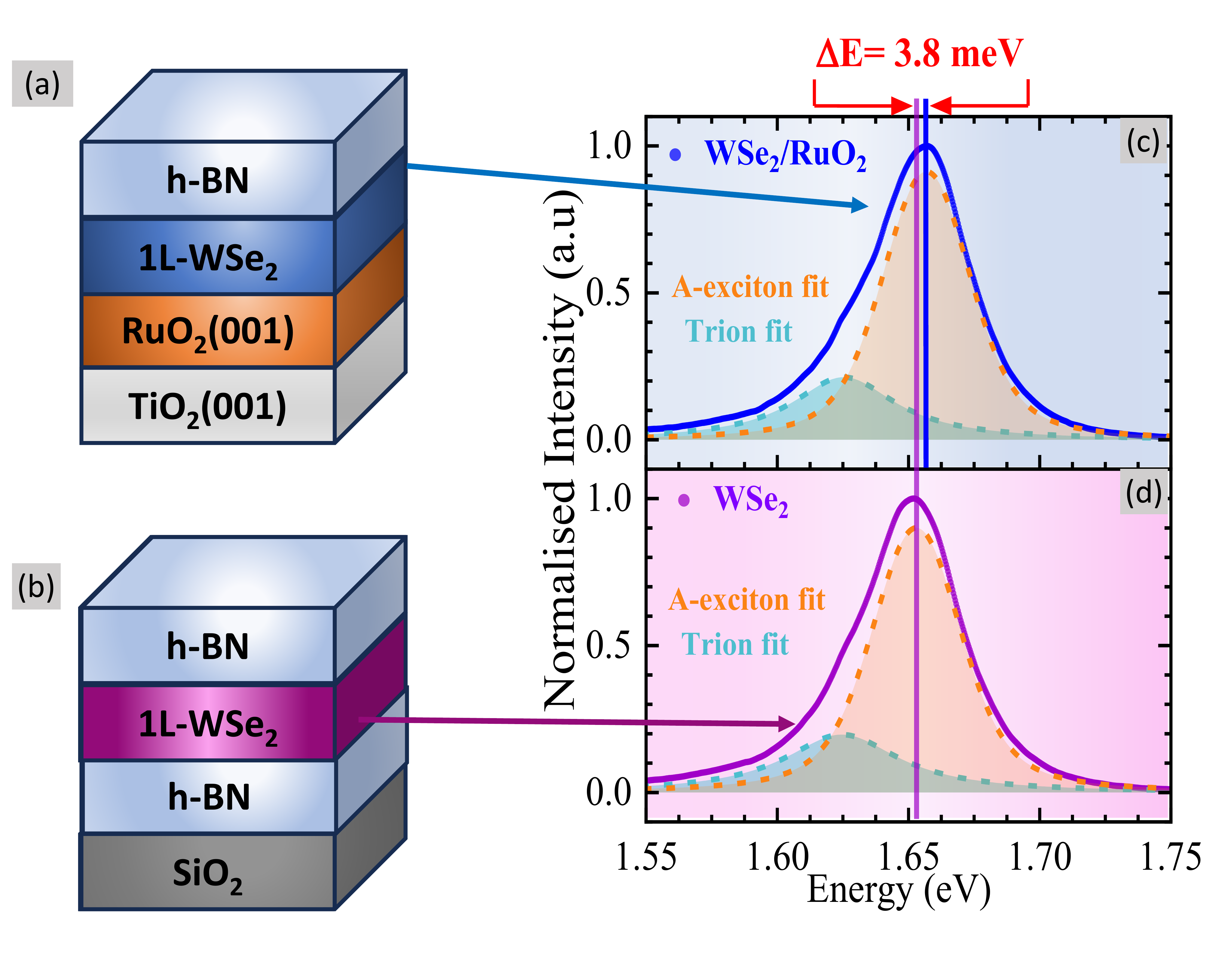}
    \caption{(a) Schematic of a heterostructure consisting of a monolayer WSe$_2$ atop a RuO$_2$ (001) film, with h-BN encapsulation on top, used for surface magnetism studies. (b) Schematic of the control sample: a monolayer WSe$_2$ encapsulated by h-BN on both top and bottom. (c,d) Room-temperature PL spectra of monolayer WSe$_2$ from the devices shown in (a) and (b), respectively. The dotted lines represent Lorentzian fits for A-excitons (yellow) and trions (cyan) used to extract peak energies. The peak energy of WSe$_2$ atop RuO$_2$ is blue-shifted by $3.8$$\pm$0.1 meV compared to the encapsulated bare WSe$_2$.}
\end{figure}
We begin by fabricating a van der Waals heterostructure of RuO$_2$ (001) and monolayer WSe$_2$ to probe the surface magnetization in RuO$_2$. To decouple the influence of RuO$_2$ on the optical response of WSe$_2$, we fabricated a control sample consisting of a monolayer WSe$_2$ encapsulated by h-BN. The schematic representations of WSe$_2$ atop RuO$_2$ (WSe$_2$/RuO$_2$) and encapsulated WSe$_2$ are shown in Figures 1(a) and 1(b), respectively. The monolayer WSe$_2$ and few-layer h-BN were obtained using mechanical exfoliation and subsequently transferred onto RuO$_2$ and SiO$_2$ using a dry transfer technique (Figures 1(a,b)) \cite{shaikh_bridging_nodate}. RuO$_2$ films were deposited on TiO$_2$ (001) substrates by reactive magnetron sputtering, with details provided in the Methods section. The thickness of the RuO$_2$ film was measured using X-ray reflectometry and found to be 54 nm with a roughness of 2 nm.

Room-temperature PL spectra of WSe$_2$/RuO$_2$ and encapsulated WSe$_2$ are shown in Figures 1(c) and 1(d), respectively. The PL spectra were fitted using a Lorentzian function to extract the peak energies of the A-exciton (Coulomb bound electron-hole pair) and trion (charged exciton) (Figures 1(c,d)). At room temperature, the peak energy of the A-exciton from WSe$_2$/RuO$_2$ is blue-shifted by 3.8$\pm$0.1 meV compared to that of encapsulated WSe$_2$.

\subsection{Temperature-dependent evidence of surface magnetic states in RuO$_2$}

\begin{figure}[!ht]
    \centering
    \includegraphics[width=0.9\textwidth]{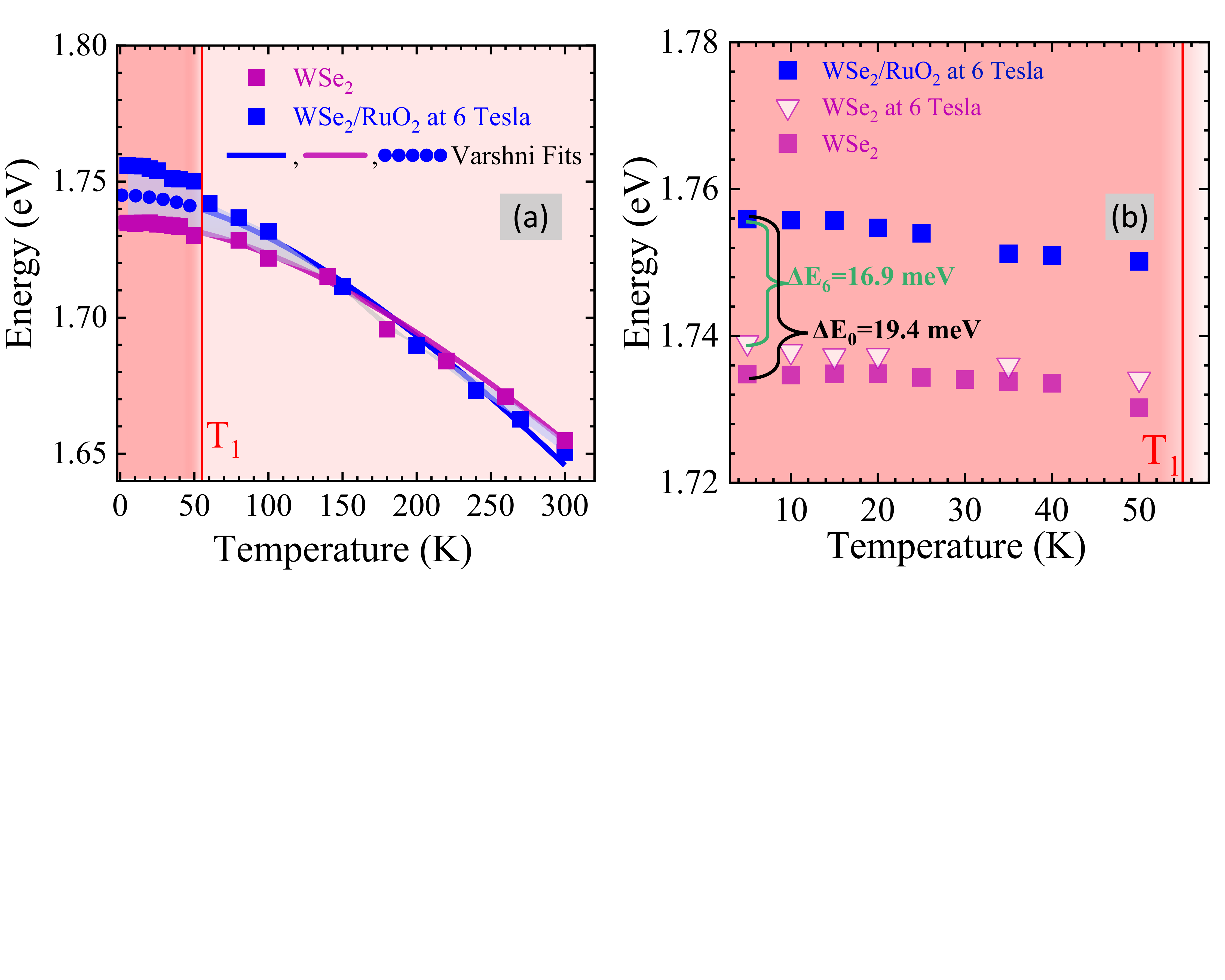}
    \vspace{-150 pt}
    \caption{(a) Peak energy of WSe$_2$ as a function of temperature. Blue squares represent the peak energy of WSe$_2$ atop RuO$_2$ in the presence of a 6 T out-of-plane magnetic field. Purple squares represent the peak energy of encapsulated WSe$_2$ without any applied magnetic field. Solid lines are Varshni fits for encapsulated WSe$_2$ (purple) and WSe$_2$ atop RuO$_2$ (blue). Encapsulated WSe$_2$ peak energy follows the Varshni fit across the entire temperature range, while WSe$_2$ atop RuO$_2$ follows it until 55 K (T$_1$). Beyond that, the Varshni behavior is shown by the blue dotted line. (b) Peak energy below T$_1$ for WSe$_2$ atop RuO$_2$ in the presence of 6 T (blue squares), encapsulated WSe$_2$ at 0 T (purple squares), and encapsulated WSe$_2$ in the presence of 6 T (purple triangles), comparing the magnitude of the energy shift of WSe$_2$ atop RuO$_2$ with that of encapsulated WSe$_2$.}
\end{figure}

To disentangle the MPE from other possible mechanisms that could contribute to the observed 3.8$\pm$0.1 meV energy shift — such as dielectric screening, charge transfer, and interfacial strain, all of which can cause band renormalization and alter the excitonic transition \cite{zhao_interlayer_2024, Raja2017-cb, PhysRevLett.113.076802}, we performed temperature-dependent PL measurements, tracking the A-excitonic peak energy shift as a function of temperature (Figure 2) \cite{zhao_interlayer_2024, Raja2017-cb, PhysRevLett.113.076802}. Previous studies have shown that the excitonic peak energy can deviate from the Varshni effect (the standard bandgap renormalization influenced by lattice expansion and electron-phonon interaction \cite{Huang2016-ud, VARSHNI1967149}) due to an underlying magnetic substrate \cite{zhao_interlayer_2024}. This deviation can also be influenced by magnetic moment orientation.

Next, we performed magnetic field cooling experiments on the RuO$_2$/WSe$_2$ heterostructure. Field-cooling through  magnetic transitions are known to saturate different magnetic domains into a single domain or to introduce preferential orientation\cite{zhao_interlayer_2024}, which can enhance the effect of small magnetic moments from RuO$_2$ on WSe$_2$. Figure 2(a) shows the peak energy of WSe$_2$/RuO$_2$ in the presence of a 6 T out-of-plane magnetic field and that of encapsulated WSe$_2$ with no magnetic field, both as a function of temperature. The encapsulated WSe$_2$ follows the Varshni fit across the entire temperature range, while WSe$_2$/RuO$_2$ follows it only up to $\sim$55 K (T$_1$) (Figure 2(a)). Below T$_1$, we observe an additional peak energy shift of 8.1$\pm$0.1 meV. To confirm whether the observed effect in WSe$_2$ originates from the applied magnetic field or the RuO$_2$ substrate, we performed the same measurement on encapsulated WSe$_2$, now in the presence of a 6 T out-of-plane magnetic field (Figure 2(b)). The encapsulated WSe$_2$ still follows the Varshni behavior, in contrast to WSe$_2$/RuO$_2$. The peak energy separation between WSe$_2$/RuO$_2$ and encapsulated WSe$_2$ at 6 T reaches 16.9$\pm$0.1 meV at low temperatures (Figure 2(b)). This value is much larger than the change in peak energy expected from dielectric screening, strain, or other non-magnetic substrate-induced effects, which are generally limited to a few meV \cite{Raja2017-cb, chakraborty_quantum-confined_2017, chakraborty_electrical_2019}. Moreover, those few meV-scale changes typically follow the Varshni effect across the entire temperature range. The sharp deviation from Varshni behavior below T$_1$, with a total change of 16.9$\pm$0.1 meV, suggests that the observed effect arises from surface magnetic states in RuO$_2$ within this temperature range. This anomalous peak energy behavior is consistent with other van der Waals magnetic heterostructures \cite{zhao_interlayer_2024}. The temperature range also agrees with reported magnetic effects in thin-film structures of RuO$_2$/NiFe and RuO$_2$/Fe \cite{abel2025probingmagneticpropertiesruo2}.

Additionally, the direction of the anomalous change in the peak energy of WSe$_2$/RuO$_2$ can be controlled by the applied out-of-plane magnetic field (Figure 3). We performed peak energy measurements on WSe$_2$/RuO$_2$ as a function of temperature from the same location in the heterostructure, in the presence of $\pm$6 T out-of-plane magnetic fields (Figure 3(a)). In both field directions, deviations from Varshni behavior emerge below T$_1$, while a pronounced field-dependent energy separation develops below approximately 33 K (T$_2$). Below T$_2$, the exciton peak energy shifts in opposite directions under $+$6 T and $-$6 T, producing a clear energy splitting between the two field polarities.

The existence of the two characteristic temperature scales (T$_1$ and T$_2$) suggests a multi-stage evolution of the interfacial magnetic response at the RuO$_2$ surface. The onset temperature T$_1$ may correspond to the emergence of fluctuating or short-range-correlated surface magnetic states that begin to influence the excitonic energy of WSe$_2$. In contrast, the stronger field-polarity-dependent splitting that develops below T$_2$ likely reflects enhanced field alignment, reduced fluctuation, or increased coherence of these interfacial moments at lower temperatures. Rather than indicating long-range magnetic order, the observed behavior is more consistent with the gradual development of weak surface-localized magnetic states with temperature-dependent anisotropy and field response.

The strong dependence of the excitonic energy on the sign of the out-of-plane magnetic field further suggests that, during field cooling through the magnetic phase transition, the interfacial magnetic moments associated with the RuO$_2$ surface can be reoriented by the applied magnetic field direction. This behavior is notable because previous studies of RuO$_2$/NiFe and RuO$_2$/Fe primarily revealed in-plane exchange-related magnetic responses \cite{abel2025probingmagneticpropertiesruo2}. In contrast, the present measurements indicate that the surface magnetic states probed by WSe$_2$ possess a substantial out-of-plane field response. Such behavior may originate from reduced symmetry and modified magnetic anisotropy at the RuO$_2$ surface, where broken inversion symmetry, oxygen non-stoichiometry, disorder-induced canting, interfacial spin-orbit coupling, or uncompensated local moments can generate weak surface magnetization components that differ substantially from the bulk magnetic response. Additionally, Ru atoms at the WSe$_2$/RuO$_2$ interface are expected to be less effectively screened than Ru atoms in the bulk; this reduced screening can increase the local effective potential U$_{\text{eff}}$, pushing the interfacial Ru sites closer to a moment-forming instability even when bulk RuO$_2$ remains nonmagnetic. The ability of the surface moments to respond to magnetic fields applied along different directions during field cooling further suggests relatively weak magnetic anisotropy and a potentially flexible interfacial spin configuration.
To further quantify the effect below T$_2$, we plot the energy difference $\Delta$E between WSe$_2$/RuO$_2$ and encapsulated WSe$_2$ under $\pm$6 T in Figure 3(b). Under $+$6 T, $\Delta$E increases with decreasing temperature and reaches 21.4$\pm$0.1 meV at 5 K, whereas under $-$6 T, $\Delta$E decreases to 14.9$\pm$0.1 meV at 5 K. The resulting field-dependent energy difference of 6.5$\pm$0.1 meV at 5 K further supports the existence of field-responsive interfacial magnetic states associated with the RuO$_2$ surface.

\begin{figure}[!ht]
    \centering    \includegraphics[width=0.9\textwidth]{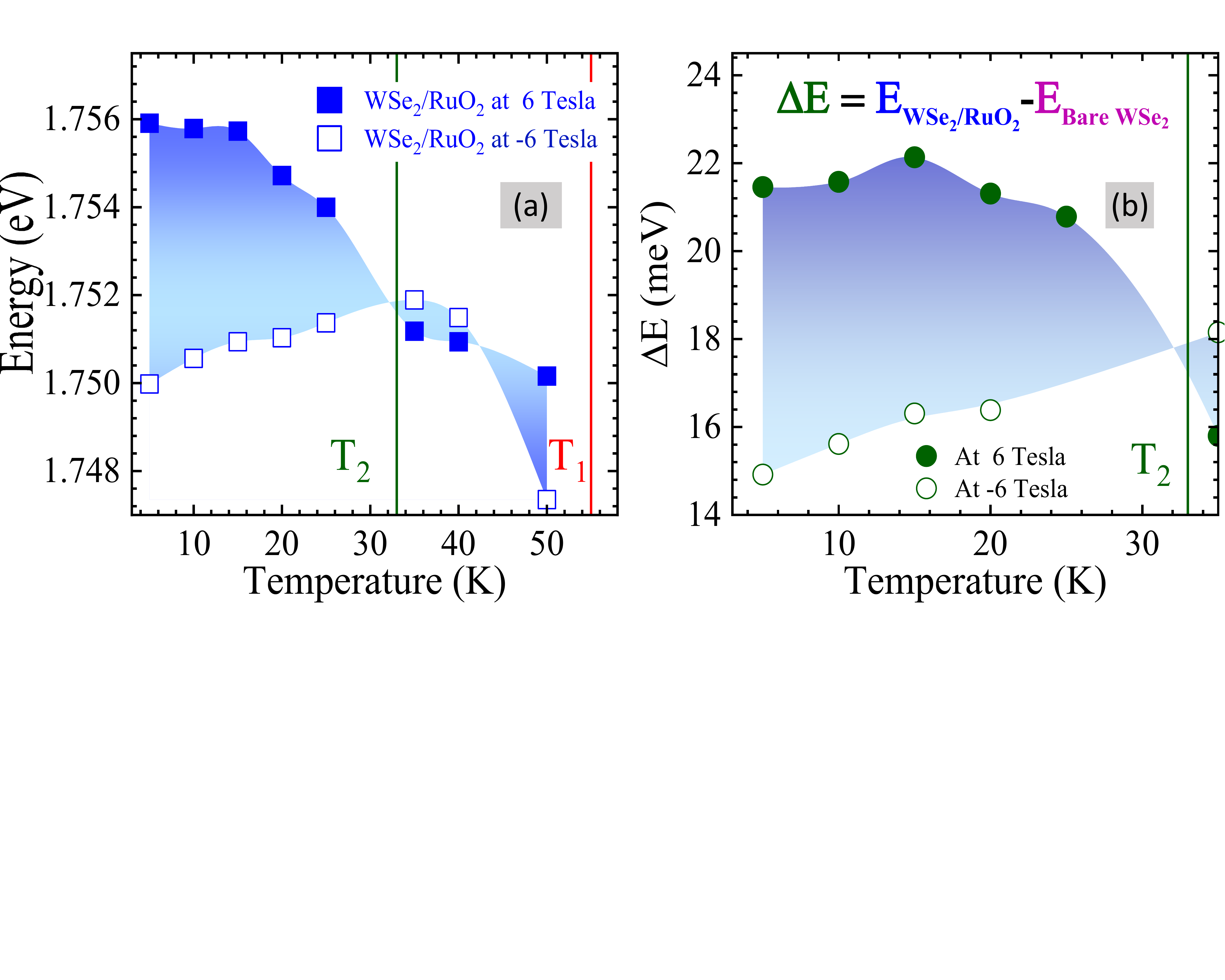}
    \vspace{-150 pt}
    \caption{(a) Peak energy of WSe$_2$ atop RuO$_2$ as a function of temperature in the presence of $\pm$6 T. The separation between the peak energies for positive and negative fields is minimal between 55 K (T$_1$) and 33 K (T$_2$), and increases below T$_2$. (b) Magnitude of the energy difference between WSe$_2$ atop RuO$_2$ and encapsulated WSe$_2$ as a function of temperature below T$_2$ in the presence of $\pm$6 T.}
\end{figure}


\subsection{Magnetic-field sweep dependent evidence of surface magnetic states in RuO\texorpdfstring{$_2$}{2}}

\begin{figure}[!ht]
    \centering
    \includegraphics[width=0.8\textwidth]{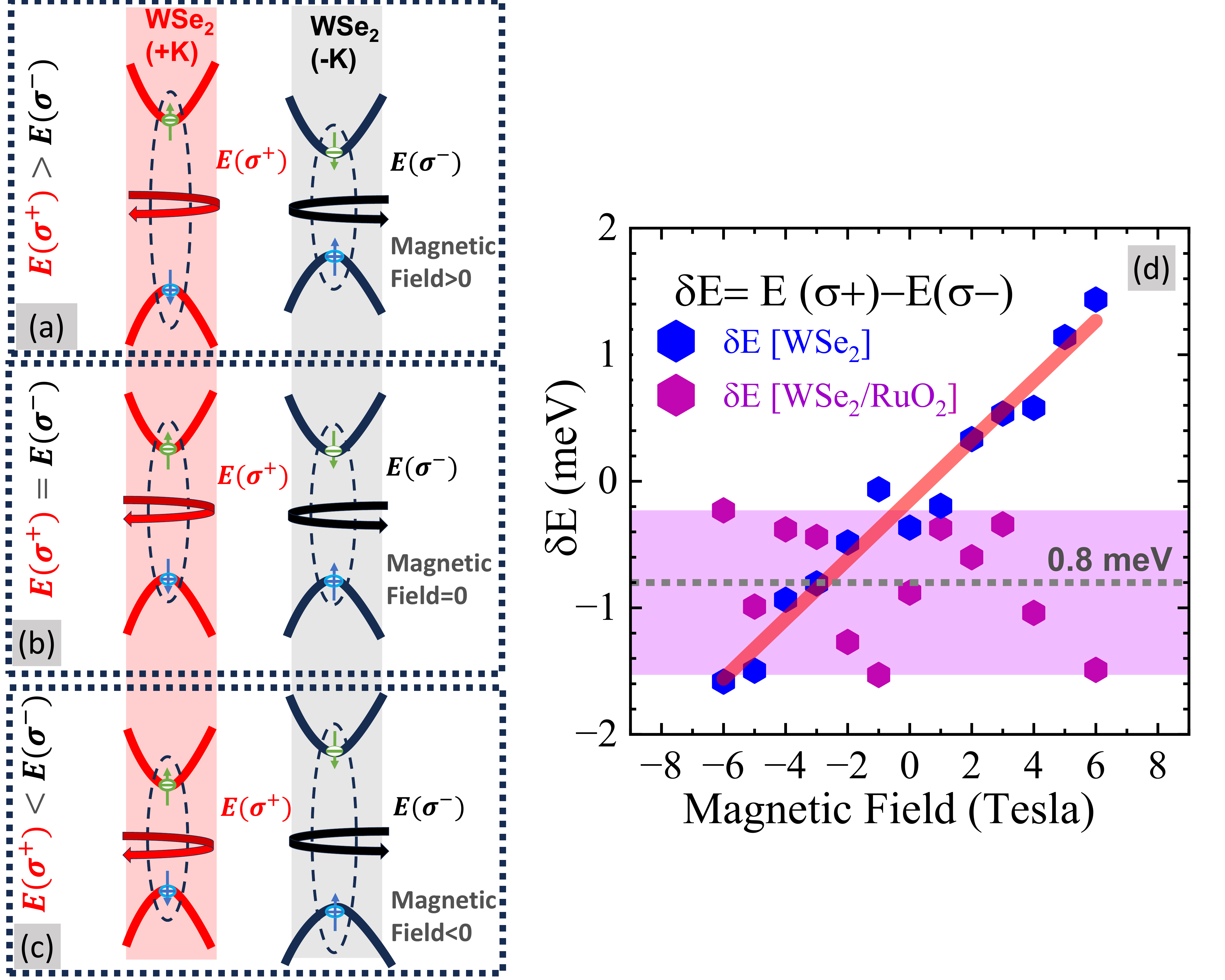}
    \caption{(a), (b), and (c) show schematics of valley-specific transitions in WSe$_2$ and a comparison between right-handed and left-handed emission energies ($E(\sigma+)$ and $E(\sigma-)$) for magnetic fields greater than, equal to, and less than zero, respectively. (d) Valley splitting energy magnitude ($\delta E$) as a function of magnetic field for encapsulated WSe$_2$ (blue diamonds) and WSe$_2$ atop RuO$_2$ (purple diamonds). The red line shows the linear fit for the Zeeman splitting behavior of encapsulated WSe$_2$, from which the extracted g-factor was found to be 4.07 $\pm$ 0.24. The highlighted region shows the minimum and maximum fluctuation in energy splitting values in the WSe$_2$/RuO$_2$ heterostructure, and the mean value of the energy fluctuation is shown by the grey dotted line. The g-factor for the WSe$_2$/RuO$_2$ heterostructure is negligible.}
\end{figure}

After observing the effect of interfacial magnetic states in the WSe$_2$/RuO$_2$ heterostructure, we next performed magnetic field sweep measurements at 1.8 K. These measurements provide insight into how the magnetic moments align themselves with respect to the applied out-of-plane magnetic field. In particular, we sought to determine whether the interfacial magnetic configuration established during field cooling remains rigid or can continue to evolve during subsequent magnetic field sweeps.

We first cooled the sample from room temperature to 1.8 K in the presence of a -6 T applied magnetic field, and then carried out the magnetic sweep measurement. To study the valley-specific transitions in WSe$_2$/RuO$_2$ and encapsulated WSe$_2$, we performed polarization-resolved spectroscopy on both samples \cite{srivastava_valley_2015}. The sample was excited with $\sigma+$ or $\sigma-$ polarized light, and the same circularly polarized emission was detected. Figures 4(a-c) show schematics of the circularly polarized transitions in WSe$_2$ for magnetic fields greater than, equal to, and less than zero, respectively. We then calculated the energy difference ($\delta E$) by subtracting the peak energies of the two circularly polarized transitions from each other. The encapsulated WSe$_2$ follows the typical linear Zeeman splitting behavior $\delta E = g \mu_B B$. The g-factor was extracted using a linear fit (red line in Figure 4(d)) and found to be 4.07 $\pm$ 0.24, consistent with previous studies \cite{srivastava_valley_2015} (Figure 4(d)). In stark contrast, the Zeeman splitting in WSe$_2$/RuO$_2$ shows no significant trend, with a g-factor close to zero. The splitting fluctuates between a minimum of -0.2 $\pm$ 0.1 meV and a maximum of -1.5 $\pm$ 0.1 meV, as shown by the highlighted region in Figure 4(d), with a mean fluctuation of 0.8 $\pm$ 0.1 meV indicated by the dotted grey line.

The absence of a conventional linear Zeeman response suggests that the valley states in WSe$_2$/RuO$_2$ are dominated by interfacial exchange fields associated with the RuO$_2$ surface rather than by the external magnetic field alone. The weak field dependence, together with the fluctuating valley splitting, further suggests that the interfacial magnetic states arise from spatially nonuniform surface moments. The change in g-factor from 4.07 $\pm$ 0.24 to negligible value supports this interpretation and rules out changes in the local electric field environment due to the different dielectric environment. The electric field environment can only change the g-factor by a small fraction ($\leq$0.5), as observed in InGaAs quantum dots \cite{prechtel_electrically_2015} and in monolayer WSe$_2$ \cite{chakraborty_quantum-confined_2017, zhou_enhanced_2025}. This interpretation is consistent with our earlier temperature-dependent measurements, which showed that the surface magnetic response can be altered by the direction of the applied magnetic field. The unsual out-of-plane field response also contrasts with the predominantly in-plane exchange behavior previously observed in RuO$_2$/NiFe and RuO$_2$/Fe heterostructures \cite{abel2025probingmagneticpropertiesruo2}, suggesting a distinct magnetic environment producing anomalous magneto-optical response  at the RuO$_2$/WSe$_2$ interface.

Possible origins of this behavior include reduced symmetry at the RuO$_2$ surface, the requirement of a high magnetic field to change the interfacial spin configuration, disorder-induced canting, oxygen-vacancy-related local moments, interfacial spin-orbit coupling, or frustrated exchange interactions that generate weak surface magnetization components. In such a scenario, the external magnetic field does not simply induce a conventional Zeeman splitting, but instead modifies the local interfacial exchange landscape experienced by the WSe$_2$ valley states.

\section{Conclusion and Outlook}

In this work, we probed the surface layer magnetism of a magnetron sputtered RuO$_2$ film using the magnetic proximity effect in a van der Waals heterostructure with monolayer WSe$_2$. Temperature-dependent magneto-optical spectroscopy revealed an anomalous excitonic energy shift and a clear deviation from conventional Varshni behavior below T$_1 \sim$ 55 K in the WSe$_2$/RuO$_2$ heterostructure, while no such behavior was observed in the encapsulated control sample. The anomalous energy shift reverses sign when the direction of the applied out-of-plane magnetic field is reversed, indicating that the associated surface magnetic moments can be reoriented by the external magnetic field.

The substantial out-of-plane field response observed in the WSe$_2$/RuO$_2$ heterostructure is particularly notable because previous studies of RuO$_2$/NiFe and RuO$_2$/Fe primarily reported in-plane exchange-related magnetic behavior. In contrast, our measurements suggest that the surface magnetic states probed by WSe$_2$ possess a sizable out-of-plane response and can be modified by the external magnetic field direction. A likely microscopic mechanism is that reduced screening and symmetry breaking at the RuO$_2$ surface enhance the local effective correlation strength on Ru atoms, allowing surface-layer moments to emerge even when the bulk remains nonmagnetic. The ability of these moments to respond to magnetic fields applied along different directions further suggests a flexible interfacial spin configuration.

It is important to emphasize that our measurements demonstrate the presence of weak interfacial magnetic states in RuO$_2$, rather than providing direct evidence of altermagnetism. Although RuO$_2$ has been theoretically proposed as an altermagnetic candidate, our optical measurements do not directly probe the momentum-dependent spin-split electronic structure characteristic of altermagnetic systems. Furthermore, many theoretical predictions of magnetic ordering in RuO$_2$ rely on specific conditions such as epitaxial strain, reduced dimensionality, a large on-site Coulomb interaction U, or crystallographic orientations different from the (001)-oriented films studied here. Our results instead demonstrate that magnetic-proximity-based optical spectroscopy using TMDC heterostructures can serve as a highly sensitive probe of weak surface and interfacial magnetic states in RuO$_2$.

More broadly, these results establish magnetic-proximity-based optical probing as a powerful approach for investigating weak surface magnetism in systems where conventional bulk magnetometry techniques may lack sufficient sensitivity. An important advantage of the present approach is that the observed magnetic proximity response is obtained without introducing an additional ferromagnetic layer. The WSe$_2$/RuO$_2$ heterostructure enables direct optical probing of emergent surface magnetic states intrinsic to the RuO$_2$ interface itself, providing a cleaner platform for investigating weak interfacial magnetism in complex oxide systems. Future studies combining optical spectroscopy with complementary transport probes may help further clarify the microscopic origin and symmetry of the observed interfacial magnetic states. In particular, angle-dependent transport and Hall measurements correlated with valley-resolved optical spectroscopy could provide important insight into the anisotropy and symmetry of the interfacial exchange coupling. Extending the measurements to different RuO$_2$ surface orientations and thicknesses may further reveal how interfacial symmetry and spin-orbit coupling influence surface magnetic behavior. Ultimately, the ability to optically detect and manipulate weak interfacial magnetic states may provide new opportunities for exploring emergent spin phenomena and future low-power spintronic functionalities.

\section{Methods}

\subsection{Preparation of heterostructure}

WSe$_2$ and h-BN bulk crystals were exfoliated using low-residue tape. The thin crystals were then transferred onto polydimethylsiloxane (PDMS). Monolayer WSe$_2$ and thin h-BN were identified using optical contrast. Monolayer WSe$_2$ was further confirmed by PL and Raman spectroscopy. The heterostructures were assembled on RuO$_2$ and SiO$_2$ substrates using a sequential dry transfer technique with PDMS \cite{shaikh_bridging_nodate}. 

\subsection{Preparation of RuO$_2$ Films}

RuO$_2$ films were deposited by reactive magnetron sputtering onto TiO$_2$ (001) substrates heated to 450 $^\circ$C. The deposition chamber had a base pressure of less than $1 \times 10^{-6}$ Torr. A 200 W DC power was applied to the Ru target in a 1:1 Ar/O$_2$ atmosphere at a total pressure of 15 mTorr. 

\subsection{Magneto-optical spectroscopy}

Temperature-dependent and polarization-dependent magneto-photoluminescence measurements were performed using an Attocube-2100 cryostat equipped with a superconducting magnet in Faraday geometry, with an applied magnetic field range of up to $\pm 9$ T. The Attocube cryostat is equipped with a confocal microscope for temperature- and polarization-resolved photoluminescence measurements as a function of magnetic field and temperature. An objective with a numerical aperture (NA) of 0.82 was used to excite the sample with a 532 nm continuous-wave laser. The spot size was approximately 791.5 nm. Circularly polarized light ($\sigma^+$ and $\sigma^-$) was produced and detected using a combination of a linear polarizer and a quarter-wave plate from Thorlabs. The emitted light was collected through the microscope's collection arm, coupled to a fiber-optic cable, and directed to the entrance port of a triple-grating Teledyne HRS-750 imaging spectrometer with a focal length of 750 mm. The collimated light was diffracted by the grating and detected by a nitrogen-cooled CCD camera with a $1340 \times 400$ pixel array, enabling precise spectral measurements.

\section{Acknowledgments}

This research was partially supported by the National Science Foundation (NSF) through the University of Delaware Materials Research Science and Engineering Center (MRSEC) Seed Award program (DMR-2011824). The authors acknowledge the use of the Materials Growth Facility (MGF) at the University of Delaware, which is partially supported by the National Science Foundation Major Research Instrumentation under Grant No.1828141 and UD-CHARM, a National Science Foundation MRSEC, under Award No. DMR-2011824. M.H.S., C.M., and C.C. acknowledge partial support from NSF Award OIA-2217786. S.B., D.T.P., M.G., and J.Q.X. acknowledge support from NSF DMR-2316664 and the King Abdullah University of Science and Technology (KAUST), ORFS-2022-CRG11-5031.2. The h-BN crystal was grown by K.W. and T.T. The authors thank Xi Wang from the Department of Materials Science and Engineering, University of Delaware, for providing access to their room-temperature PL and Raman measurement setup. The authors also thank Yi Ji from the Department of Physics and Astronomy, University of Delaware, for providing access to their optical microscope.

\section{Data Availability Statement}
The data that support the findings of this study are available from the corresponding author upon reasonable request.}


%
\bibliographystyle{MSP}
\bibliography{reference}

\end{document}